# *PT*-symmetry enabled spin circular photogalvanic effect in antiferromagnetic insulators

Ruixiang Fei,[1,†] Wenshen Song,[1] Lauren Pusey-Nazzaro,[1] Li Yang [1, 2, ‡]

[1] *Department of Physics, Washington University in St Louis, St Louis, Missouri 63130, United States*

[2] *Institute of Materials Science and Engineering, Washington University in St. Louis, St. Louis, Missouri 63130, United States*

**Abstract**

The short timescale spin dynamics in antiferromagnets is an attractive feature from the standpoint of ultrafast spintronics. Yet generating highly polarized spin currents at room temperature remains a fundamental challenge for antiferromagnets. We propose a spin circular photogalvanic effect (spin-CPGE), in which circularly polarized light can produce a spin current without net charge current at room temperature, through an "injection-current-like" mechanism in parity-time(*PT*)-symmetric antiferromagnetic (AFM) insulators. We demonstrate this effect by first-principles simulations of bilayer $CrI_3$ and room-temperature AFM hematite. Our calculations show that the spin-CPGE is significant, and the magnitude of spin photo-current is comparable with the widely observed charge photocurrent in ferroelectric materials. Interestingly, this spin photocurrent is not sensitive to spin-orbit interactions, which were regarded as fundamental mechanisms for generating spin current. Given the fast response of light-matter interactions, large energy scale, and insensitivity to spin-orbit interactions, our work gives hope to realizing a fast-dynamic and temperature-robust pure spin current in a wide range of *PT*-symmetric AFM materials, including weak-relativistic magnetic insulators and topological axion insulators.

**Introduction**: Spintronics based on antiferromagnetic (AFM) materials has great potential in reducing device scale and power consumption [1–3]. In contrast to ferromagnets, antiferromagnets are robust against perturbed magnetic fields, produce no stray fields, and are capable of generating large magneto-transport effects [1], making them promising for next-generation spintronics. Particularly, antiferromagnets exhibit unique advantages in generating spin current, which is the basis for spintronic devices. Compared with gigahertz microwave pulses used in ferromagnets, terahertz pulses can pump spin current by exciting the left- and right-hand mode of magnons in antiferromagnets [4–6] due to the strong exchange interaction, giving rise to ultrafast spin dynamics [7]. Besides, spin caloritronics based on magnons in AFM materials, such as the spin Seebeck and spin Nernst effects [8–11], was proposed to generate pure spin current in the same device. However, because of the intrinsically small energy scales, magnon-based spin currents decay rapidly with increasing temperature [5,12], making it difficult for device applications at room temperature.

Optical pump-induced spin dynamics may be a promising approach to surmounting such difficulties because of its intrinsically larger energy scale [13,14] and observed strong light-matter interactions in magnetic materials [15–17]. Notably, second-order light-matter interactions, such as the circular photogalvanic effect (CPGE) or bulk photovoltaic effect (BPVE), are known to create charge current in polar materials at room temperature without bias [18–23]. Recently, This idea has been expanded to two-dimensional (2D) party-time (*PT*) symmetric AFM insulators [24] and topological axion insulators [25,26], in which a sizeable DC charge current was predicted to be generated by linearly polarized light [24,25,27].

Beyond charge current, light can also drive spin current. For example, spin photocurrent was predicted in AFM hematite [28] and bilayer $CrCl_3$ [29], via the shift-current mechanism under linearly polarized light, denoted as the spin-BPVE [28]. Unfortunately, linearly-polarized light simultaneously excites significant charge current because of the broken SU(2) symmetry due to spin-orbit coupling (SOC) [24,25], substantially diluting spin polarization of the overall current. It remains a fundamental challenge to generate a highly polarized spin current in AFM materials at room temperature.

In this work, we predict a spin circular photogalvanic effect (spin-CPGE), in which circularly polarized light can generate pure spin current in *PT*-symmetric antiferromagnets. This effect arises from a second-order light-matter interaction, i.e., an inject-current-like mechanism. The net charge current is zero, being a *PT*-even contribution and having a trivial non-abelian Berry connection, while the spin current is nonzero since it is a *PT*-odd contribution. We further demonstrate this effect by first-principles simulations of two typical *PT*-symmetric antiferromagnets: bilayer $CrI_3$ and three-dimensional (3D) hematite. Our result confirms enhanced pure spin current and unexpectedly shows that SOC is inessential for generating spin current in *PT*-symmetric antiferromagnets. Our work suggests that collinear *PT*-symmetric AFM insulators, even weak-relativistic AFM insulators, can serve as effective and temperature-robust spin generators for both spin orientations via ultrafast light-matter interactions.

**PT-symmetry induced chiral spin photocurrent**

Let us consider an insulating AFM crystal that breaks both time-reversal symmetry (*T*) and inversion symmetry (*P*) but respects their combination *PT*. In such a case, the Hamiltonian satisfies $H(k,\uparrow) = H(k,\downarrow)$ for any $k$ wavevector of reciprocal space, resulting in exactly double spin-degenerate bands, as shown in Figure 1(a). Because of these degenerate bands and zero Abelian Berry curvature, very subtle longitudinal and transverse spin current can be observed in transport measurements. Nevertheless, after involving the second-order light-electron interaction for the non-biased case, we will show that these degenerate spins exhibit different behaviors, enabling a desired spin-polarized current.

For a coherent light illumination, the general form of a second-order nonlinear DC photocurrent is $J_c = \chi_{abc}(0;\omega,-\omega)E_a(\omega)E_b(-\omega)$, where $a$ and $b$ the polarization directions of light, $c$ is the outgoing current direction, and $\chi_{abc}(0;\omega,-\omega)$ is the DC photoconductivity in the $c$-direction. For circularly polarized light, the inject-current-like photoconductivity $\eta^{\alpha}_{abc} \equiv \chi_{abc}(0;\omega,-\omega)$ under the relaxation time approximation is [19,30]

$$\eta^{\alpha}_{abc} = \frac{-\pi e^3}{\hbar^2 \omega^2} \sum_{mn} \int d^3k \, Im(v^a_{mn,\alpha}(k)v^b_{nm,\alpha}(k)) f_{mn} (\langle m|\{\sigma_\alpha, v_c\}|m\rangle \tau_m - \langle n|\{\sigma_\alpha, v_c\}|n\rangle \tau_n) \delta(\omega - \omega_{mn}), \quad (1)$$

where $Im(v^a_{mn,\alpha}(k)v^b_{nm,\alpha}(k)) = \frac{1}{2}\left(v^a_{mn,\alpha}v^b_{nm,\alpha}(k) - v^b_{mn,\alpha}(k)v^a_{nm,\alpha}(k)\right)$ is the imaginary part of optical oscillator strength. $v^a_{mn,\alpha}(k)$ and $v^b_{nm,\alpha}(k)$ are the $a$-direction and $b$-direction interband velocity matrix elements between the $m$-th and $n$-th bands with $\alpha$ spin at the same wave vector k, respectively. $f_{mn} = f(\epsilon_{mk}) - f(\epsilon_{nk})$ is the occupation number difference between the $m$-th and $n$-th bands. $\tau_m$ and $\tau_n$ are the minimum value of spin-relaxation time and free-carrier relaxation time of the $m$-th and $n$-th bands, respectively. We define $\{\sigma_\alpha, v_c\} = \frac{1}{2}(v_c \sigma_\alpha + \sigma_\alpha v_c)$, whence $\langle m|\{\sigma_\alpha, v_c\}|m\rangle$ is the $c$-direction velocity matrix of the $m$-th band electron with the $\alpha$-direction spin.

From *PT*-symmetry, it follows that the electronic bands are spin degenerate, and the group velocity matrix of two spins should be in the same direction, namely, the spin velocity matrix satisfy $\{\sigma_\uparrow, v_c\} = \{\sigma_\downarrow, v_c\}$. However, the *PT*-symmetry enforces a constraint on the imaginary part of the optical oscillator strength: $Im(v^a_{mn,\uparrow}(k)v^b_{nm,\uparrow}(k)) = -Im(v^a_{mn,\downarrow}(k)v^b_{nm,\downarrow}(k))$. Therefore, the spin-"up" photoconductivity of Eq. 1 is opposite to the spin-"down" photoconductivity at the same *k* point. Since the overall spin photoconductivity is defined as $\eta^s = \eta(\uparrow) - \eta(\downarrow)$, we expect a net spin contribution without charge contribution at each *k* point. Finally, the overall spin-current is obtained by an integral over the whole reciprocal space. As shown in Figure 1(b), if the energy contour of light-pumped free carriers is not symmetric, which is the typical case in non-centrosymmetric materials, the integral is nonzero, as is the overall spin current. Intuitively, the light-driven electron transportations are helical channels where the electron spin projection is connected with its transport direction, i. e., electrons characterized by spin "up" is traveling in one direction, while electrons characterized by spin "down" is moving in the opposite direction, generating a pure spin current shown in Figure 1 (c).

It is worth noting that light will create electrons and holes simultaneously, and total spin current is composited by the hole and electron quasiparticle contributions in Eq. 1 since the summation includes both conduction and valance bands. We plot Figure 1(d) to further illustrate the roles of electrons and holes and how to measure the spin current. Under circularly polarized light, the excited holes and electrons with the same spin direction should travel in the opposite direction because of their opposite transport direction. As shown in Figure 1 (d), the "spin-up" electrons and "spin-down" holes are accumulated at the left boundary of the AFM insulator. If attached to a ferromagnetic material, in which the majority is assumed to be spin "up", the spin "down" holes will annihilate with the minority spin "down" electrons in the ferromagnetic material while the spin-up electrons remain. This ultimately results in a measured spin-polarized current.

We must also remark that there are mainly two typical second-order light-matter interactions: the inject-current mechanism and the shift-current mechanism. In addition to the injection current discussed above, shift current can be excited by the circularly polarized light simultaneously. However, because the magnitude of shift current is usually two orders smaller than that of injection current [24,25,31], the overall observed photocurrent is dominated by injection current in this work.

**Large photo-driven spin current in two-dimensional materials.**

In the following, we employ first-principles simulations (see supplementary information for simulation details [32]) [33–35] to quantitatively calculate the spin photocurrent of *PT*-symmetric AFM insulators based on the aforementioned injection-current mechanism. We first choose bilayer AFM CrI$_3$ as the prototype 2D material. The top view of bilayer CrI$_3$ is illustrated by Figure 2(a). To address the essential hexagonal structurem, we only plot the Cr atoms. The blue and red atoms represent the spin-up-layer and spin-down-layer Cr atoms, respectively. Our first-principles calculations indicate that the ground state of bilayer CrI$_3$ has an interlayer AFM order, in agreement with previous theoretical and experimental results [17,36,37]. Such a ground-state structure preserves *PT*-symmetry, so that bilayer CrI$_3$ has doubly degenerate bands at any k-point in momentum space, as confirmed by our first-principles density functional theory (DFT) result in Figure 2(b).

Figure 2(c) shows the calculated spin photoconductivity according to Eq. 1. We set the polarization of the circularly polarized light to be in the *xy*-plane and measure the $S_z$ component of spin current. The grey solid line and dashed line show the spin current along both x-direction ($\eta_x^{S_z}$) and y-direction ($\eta_y^{S_z}$), respectively. It is worth mentioning that the quantitative photoconductivity needs an estimated carrier/spin lifetime $\tau$ (Eq. 1). There is no widely accepted value of the carrier/spin lifetime of CrI$_3$. Previous works [24] adopted a value of ~0.4 ps, which is smaller than those of typical transition metal dichalcogenides (TMDs), such as MoS$_2$ (~ 1ps) [38]. In this work, we choose a more conservative value of 0.1 ps.

This spin photoconductivity is significant and expected to be detected by experiments in near future. As shown by the dashed line spectrum in Figure 2(c), the magnitude of spin photoconductivity ($\eta_y^{Sz}$) can reach up to 40 µA/V². This is roughly one order larger than the widely observed charge photoconductivity of $BaTiO_3$ [39] and 2D GeS and its analogs [40] due to BPVE and is the same order as that of magnetic injection charge-current by linearly polarized light in bilayer $CrI_3$ [24]. Moreover, the magnitude of this intrinsic spin photocurrent is comparable with those of observed photocurrents in CdSe and GaAs quantum well structures [41–43].

Finally, we find that the direction of spin photocurrent can be switched by the Neel vector, as shown in Figure 2(d). Although both configurations in Figure 2(d) are AFM, their spin currents are opposite to each other due to opposite Neel vectors. This is because the magnetic atomic structure indicated in the blue dashed square (denoted as the "up" Neel vector) can be translated to that in the red dashed square by the time-reversal symmetry operator. Accordingly, the direction of the current should be switched by time-reversal symmetry. This correlation between the Neel vector direction and spin photocurrent might be useful for detecting the Neel vector of AFM materials, which has been regarded as a challenge for years [44].

**Sizeable photo-driven spin current without SOC.**

SOC has been viewed as the fundamental mechanism for the spin-current because it can generate spin-polarized currents from charge currents, characterized as the spin Hall effect [45–47]. Besides, the SOC lies at the heart of the photo-driven charge current in *PT*-symmetric AFM materials, e.g. bilayer $CrI_3$ [24] and even-septuple layer $MnBi_2Te_4$ [25], to break the SU(2) spin-rotation symmetry. Therefore, it is necessary to reveal the origin of such an enhanced spin photocurrent in *PT*-symmetric AFM insulators and the role of SOC.

We have calculated the spin photoconductivity $\eta$ of bilayer AFM CrI$_3$ by artificially reducing the spin-orbital interaction strength ($\lambda_{so}$). We take the *y*-direction spin photocurrent as an example (see *x*-direction spin photocurrent in supplementary information [32]), and the results are presented in Figure 3(a): the solid line is the spin photoconductivity with the intrinsic SOC strength of CrI$_3$, and the dashed line is that with the negligible SOC ($\lambda = 0.001\lambda_{so}$). The corresponding DFT-calculated band structures are provided in the supplementary information. Despite the differences from the changes of band structures due to SOC, the magnitudes of these two spectra are similar; the spin photoconductivity of the dashed line can reach 45 $\mu A/V^2$, which is slightly larger than that of spin photoconductivity with full SOC. This surprisingly indicates that SOC is not necessarily responsible for the enhanced spin photocurrent.

To further understand this result, we show the spin-texture of the valence band $E = -0.17eV$ (with zero energy set to be the top of valence bands) with full SOC for bilayer CrI$_3$ in Figure 3(b). Importantly, the bands are not symmetrical in reciprocal space, e.g. the oval-shaped energy contour of the inner band, a consequence of broken SU(2) spin-rotation symmetry by the SOC. Therefore, the spin photoconductivity at the *k* point should not cancel with that at the -*k* point. We illustrate this by plotting the distribution of spin photoconductivity in reciprocal space, shown in Figure 3(c). There is a non-odd-parity symmetry of this distribution, which results in a sizable non-zero spin current after integration.

On the other hand, after reducing the SOC ($\lambda = 0.001\lambda_{so}$), we observe a different picture: the band structure with negligible SOC is symmetrical in momentum space because of the preserved SU(2) symmetry, as shown in Figure 3(d). However, the spin texture does not exhibit any symmetry in reciprocal space. Therefore, the spin current contributed by the *k* and -*k* point in the momentum space still cannot cancel each other, although the bands are symmetric, resulting in a non-zero spin current. Figure 3(e) confirms this non-odd-parity symmetry of the spin photoconductivity distribution for photon energy at 1.2 eV.

**Large photo-driven spin current in 3D materials.**

We take bulk $\alpha$-$Fe_2O_3$, i.e. hematite, as a 3D example to demonstrate the predicted spin photocurrent, in which long-distance spin transport has been reported [48]. Bulk $\alpha$-$Fe_2O_3$ is the most stable form of iron oxides, exhibiting an AFM order with the space group $R\bar{3}c$. The magnetic moments of adjacent $Fe^{3+}$ layers couple in an antiparallel fashion, forming an overall AFM ordering [49]. Below the Morin temperature ($T_M \approx 263$ K), the direction of the magnetic moments is parallel to the z-axis [50] as shown in Figure 4(a). Hematite below $T_M$ preserve the *PT* symmetry, where the symmetry center is labeled as a dashed circle in Figure 4(a). Therefore, it is an excellent candidate to study the photo-driven spin current in weak-relativistic systems at room temperature. Although spin photocurrent under linearly polarized light was predicted in hematite by considering the shift-current mechanism [28], a simultaneously excited and strong charge current substantially reduces the spin polarization. In the following, we will show that circularly polarized light can excite enhanced spin current without charge motion via the inject-current-like mechanism predicted in this work.

We have calculated the photoconductivity $\eta$ of the spin current according to Eq. 1 by using a conservative relaxation time of $\tau = 0.1$ ps. Figure 4(b) shows the $S_z$-component spin current when the polarization of the circularly polarized light is set in the *yz*-plane, and Figure 4(c) is that generated by the *xy*-plane-polarized circularly polarized light. Interestingly, the magnitude of spin photoconductivity $\eta$ is in the same order or even higher than that of the aforementioned large SOC material, e.g. bilayer $CrI_3$. For example, as shown by the dot-dashed line spectrum in Figure 4(b), the magnitude of photoconductivity ($\eta_y^{Sz}$) can reach up to 200 μA/V$^2$ by using $\tau = 0.1$ ps in Eq. 1. Such sizeable spin photoconductivity is about one order larger than that of the charge photoconductivity of BaTiO$_3$ due to BPVE [39], and the same order as that of charge photoconductivity because of CPGE in polar insulators, e.g. GeSe [31] and CdSe [43]. Figure 4(d) shows the spin photoconductivity of y-direction current under a *xy*-plane circularly polarized light that is distributed in $\Gamma TU$ plane of the reciprocal space (see the plane in supplementary information [32] ). Similar to Figure 3(e) for AFM CrI$_3$, we do not observe any symmetry of the spin-conductivity in the momentum space, leading to such a non-zero spin photocurrent.

**Summary and outlook.**

In this work, we have demonstrated a spin-CPGE effect for generating enhanced spin photocurrent in *PT*-symmetric AFM insulators. We demonstrate that such spin current is protected by *PT* symmetry and is not sensitive to SOC because of the non-odd parity of spin current in the momentum space. We quantitatively show the generation of nonlinear DC spin current under circularly polarized light using quantum perturbation theory and first-principles calculations. Using prototypical 2D AFM insulator bilayer, CrI$_3$, and 3D weak-relativistic AFM material $\alpha$-Fe$_2$O$_3$, we predict that circularly polarized light can excite pure spin current with the same order of magnitude as the widely observed charge photocurrent in nonmagnetic materials. Such predicted spin-CPGE is accessible in many traditional and emerging AFM insulators. Among them are NiO [51] and Cr$_2$O$_3$ [52], which have been used to generate magnon spin current [5], emerging 2D AFM insulators, such as the MnPS$_3$ family materials [8,16], and topological axion insulators, such as the even-layer Mn$_2$Bi$_2$Te$_4$ family materials [53,54]. Our prediction will not only be helpful to understand recent important measurements of vertical-direction photocurrent or spin photo-current in $CrI_3$ junction devices [55], but also build a general framework to search for efficient spin-pumping via light-matter interactions at room temperature.

**ACKNOWLEDGMENT**

R.F. thank Zhiqiang Bao for fruitful discussions. R. F. and L.Y. are supported by the Air Force Office of Scientific Research (AFOSR) grant No. FA9550-20-1-0255 and the National Science Foundation (NSF) CAREER grant No. DMR-1455346. The computational resources are provided by the Stampede of Teragrid at the Texas Advanced Computing Center (TACC) through XSEDE.

Email:  [†]ruixiangfei@gmail.com
          [‡]lyang@physics.wustl.edu

**Figures**

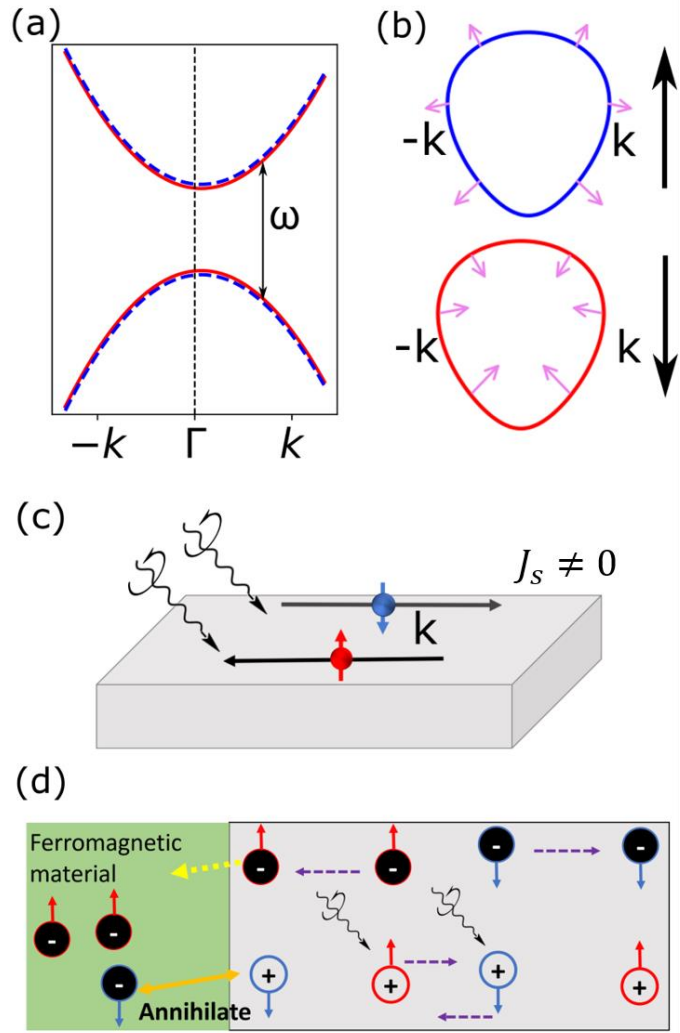

**Figure 1**. Circularly polarized light-driven spin current in an AFM insulator. (a) Doubly degenerate bands are due to *PT*-symmetry. (b) The transport direction of two opposite directions of spins after pumping. The blue and red ovals represent the contour plot of two spin components of conduction bands, and the black arrows show the overall spin current direction. (c) The schematic of circularly polarized light-induced spin current under circularly polarized light. (d) The proposed experimental setup for measuring both electron and hole spin currents. Solid and open circles represent electrons and holes, respectively.

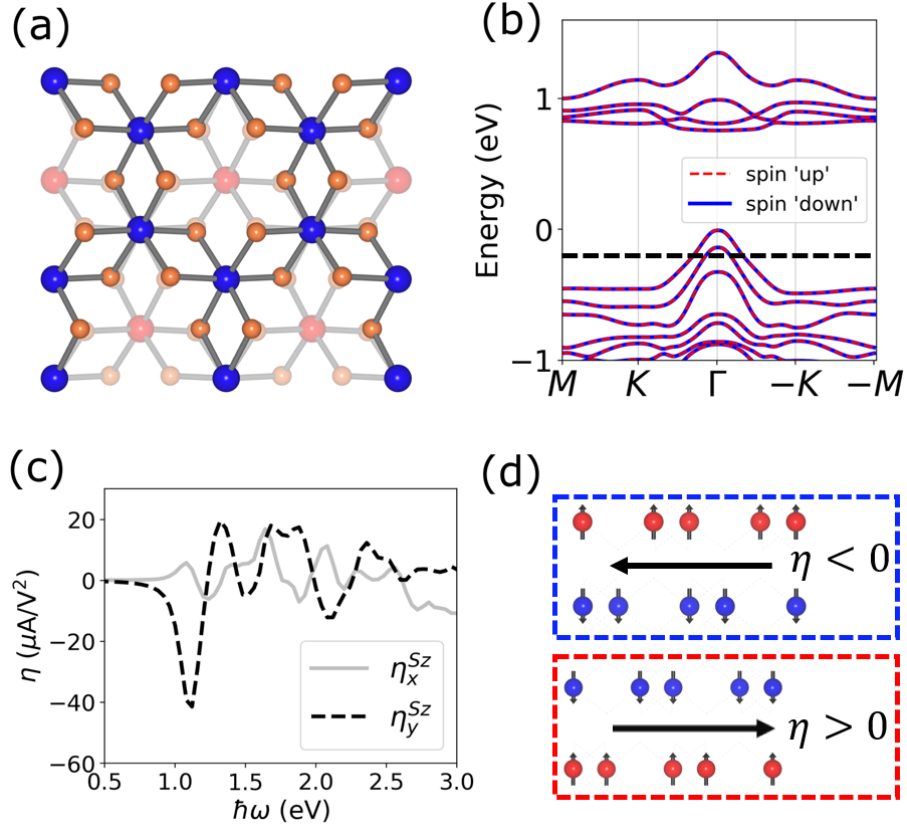

**Figure 2**. Spin CPGE in bilayer CrI$_3$. The simplified atomic structure (a) and DFT band structure (b) of bilayer AFM CrI$_3$. (c) The photo-driven spin current conductivity along the in-plane *x* and *y*-direction for the spin component of S*z*. (d) The spin current direction (black arrows) for AFM bilayer CrI$_3$ (side view of Cr atoms) with opposite Neel vector directions.

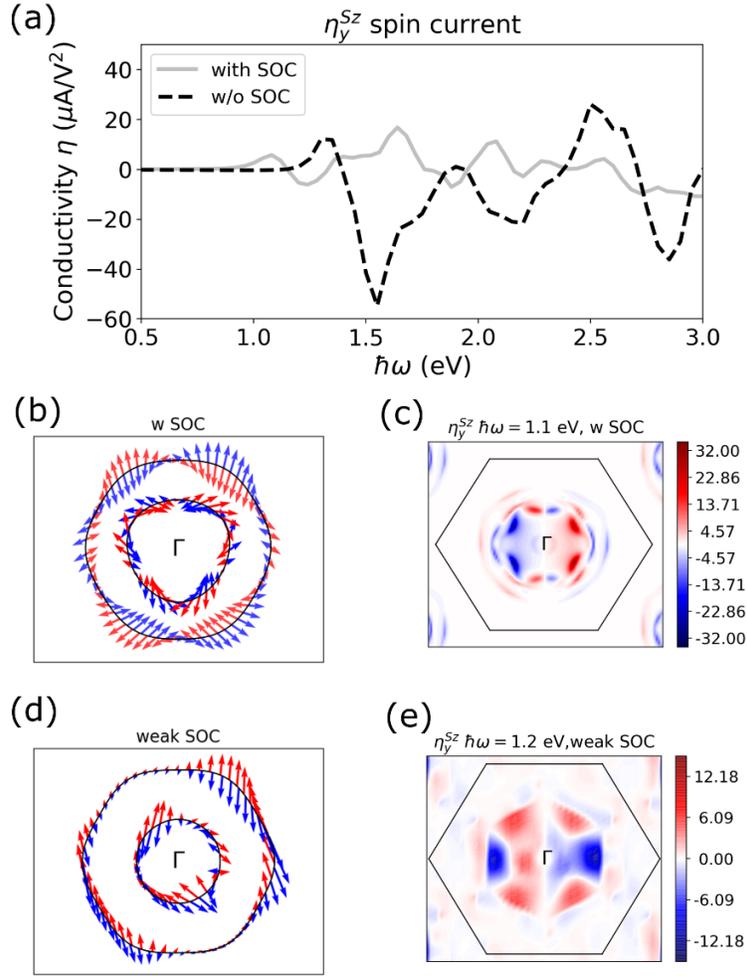

**Figure 3**. Spin photocurrent of bilayer CrI$_3$ without SOC. (a)The *y*-direction photo-driven spin current of S$z$ component with full SOC (grey solid line) and extremely weak SOC (dashed line), i.e. the SOC strength $\lambda = 0.001\lambda_{so}$. The spin texture of valence bands at $E = -0.17 eV$ in Figure 2(b) with full SOC (b) and extremely weak SOC (d). The spin photo-conductivity distribution in the momentum space with full SOC (c) and extremely weak SOC (e).

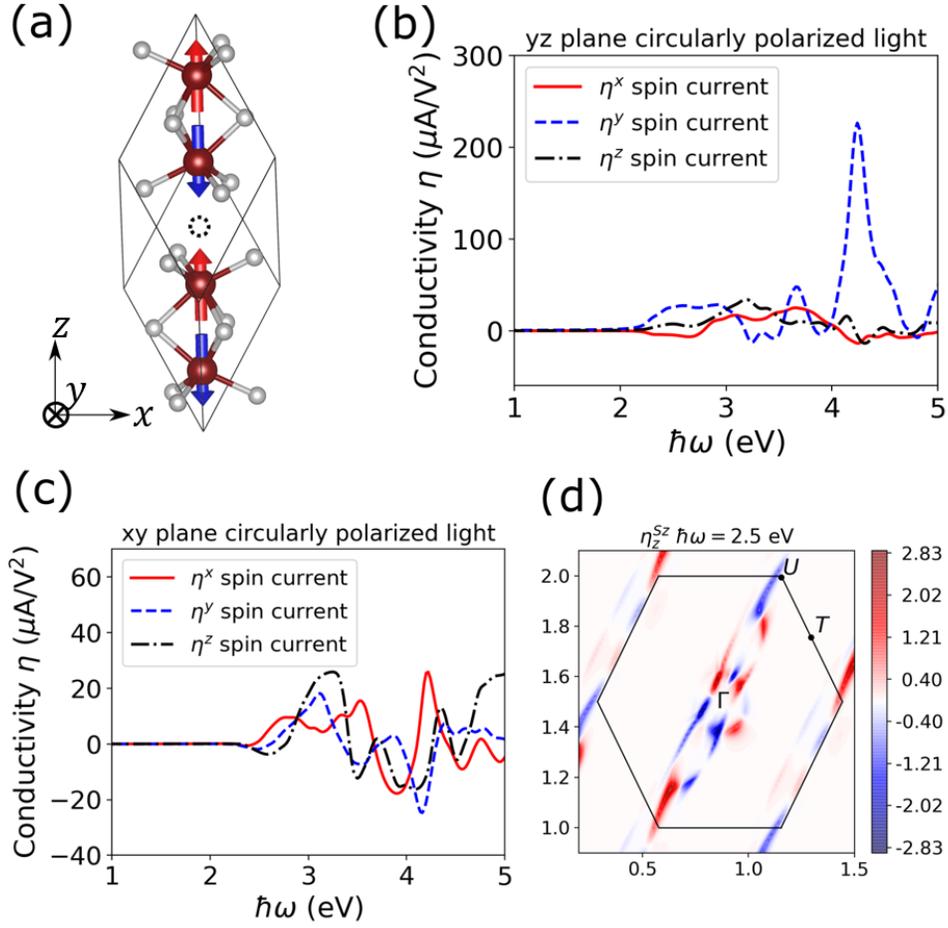

**Figure 4**. Spin CPGE in a 3D material. (a) The atomic structure of bulk $\alpha$-$Fe_2O_3$, the dashed circle being the *PT*-symmetry center. The spin photo-conductivity for the polarization of circularly polarized light in *yz* plane (b) and in *xy* plane (c). The spin photo-conductivity distribution in the momentum space (d).